\documentclass[runningheads]{llncs}
\usepackage[T1]{fontenc}
\usepackage{graphicx}

\usepackage{makecell}
\usepackage{breqn}
\usepackage[most]{tcolorbox}
\usepackage{subcaption} 
\usepackage{tabularx}
\usepackage{hyperref}
\usepackage{multirow}%
\usepackage{amsmath,amssymb,amsfonts}%
\usepackage{mathrsfs}%
\usepackage[title]{appendix}%
\usepackage{xcolor}%
\usepackage{textcomp}%
\usepackage{manyfoot}%
\usepackage{booktabs}%
\usepackage{algorithm}%
\usepackage{algorithmicx}%
\usepackage{algpseudocode}%
\usepackage{listings}%
\usepackage{wrapfig}
\begin{document}
\title{DSEBench: A Test Collection for Explainable Dataset Search with Examples}
%
%
\author{Qing Shi\thanks{These authors contributed equally.} \and
Jing He\textsuperscript{$\star$} \and
Qiaosheng Chen \and
Gong Cheng}
%
\authorrunning{Q. Shi et al.}
%
\institute{State Key Lab for Novel Software Technology, Nanjing University, Nanjing, China \\
\email{\{qingshi,522024330021,qschen\}@smail.nju.edu.cn,gcheng@nju.edu.cn}}
\maketitle              
\begin{abstract}
Dataset search is a well-established task in the Semantic Web and information retrieval research. Current approaches retrieve datasets either based on keyword queries or by identifying datasets similar to a given target dataset. These paradigms fail when the information need involves both keywords and target datasets. To address this gap, we investigate a generalized task, Dataset Search with Examples (DSE), and extend it to Explainable DSE (ExDSE), which further requires identifying relevant fields of the retrieved datasets. We construct DSEBench, the first test collection that provides high-quality dataset-level and field-level annotations to support the evaluation of DSE and ExDSE, respectively. In addition, we employ a large language model to generate extensive annotations for training purposes. We establish comprehensive baselines on DSEBench by adapting and evaluating a variety of lexical, dense, and LLM-based retrieval, reranking, and explanation methods.

\end{abstract}

\noindent\textbf{Resource type:} Datasets and annotated corpora

\noindent\textbf{License:} Apache License 2.0

\noindent\textbf{DOI:} 10.5281/zenodo.17805089

\noindent\textbf{URL:} \url{https://doi.org/10.5281/zenodo.17805089}

\section{Introduction}

Finding datasets efficiently is critical in the era of open data~\cite{DBLP:journals/vldb/ChapmanSKKIKG20,DBLP:journals/csur/PatonCW24},
and the Semantic Web community has made significant strides in this direction~\cite{DBLP:conf/semweb/ChenHZLLSC23,google_relationships,dunks}.

\textbf{Motivation.} Existing research follows two paradigms~\cite{DBLP:journals/csur/PatonCW24}: keyword-based search~\cite{GoogleDatasetSearch,DBLP:journals/kbs/SilvaB24,miu_DS,dunks} which retrieves datasets \emph{relevant to a query}, or similarity-based discovery~\cite{Auctus,TUS,D3L} which finds datasets \emph{similar to a target dataset}. Neither approach fully captures the scenarios in which users possess \emph{both} an information need expressed as a query and a few target datasets as examples. Moreover, while explainable information retrieval (IR) is trending~\cite{DBLP:conf/ecir/AnandSV25}, \emph{explainability in dataset search} remains underexplored. Current systems provide insufficient justifications,
leaving users without insight into the factors driving retrieval decisions~\cite{DBLP:journals/is/BarcellosBV22}. 

\begin{figure}[t]
\centering
\includegraphics[width=0.9\linewidth]{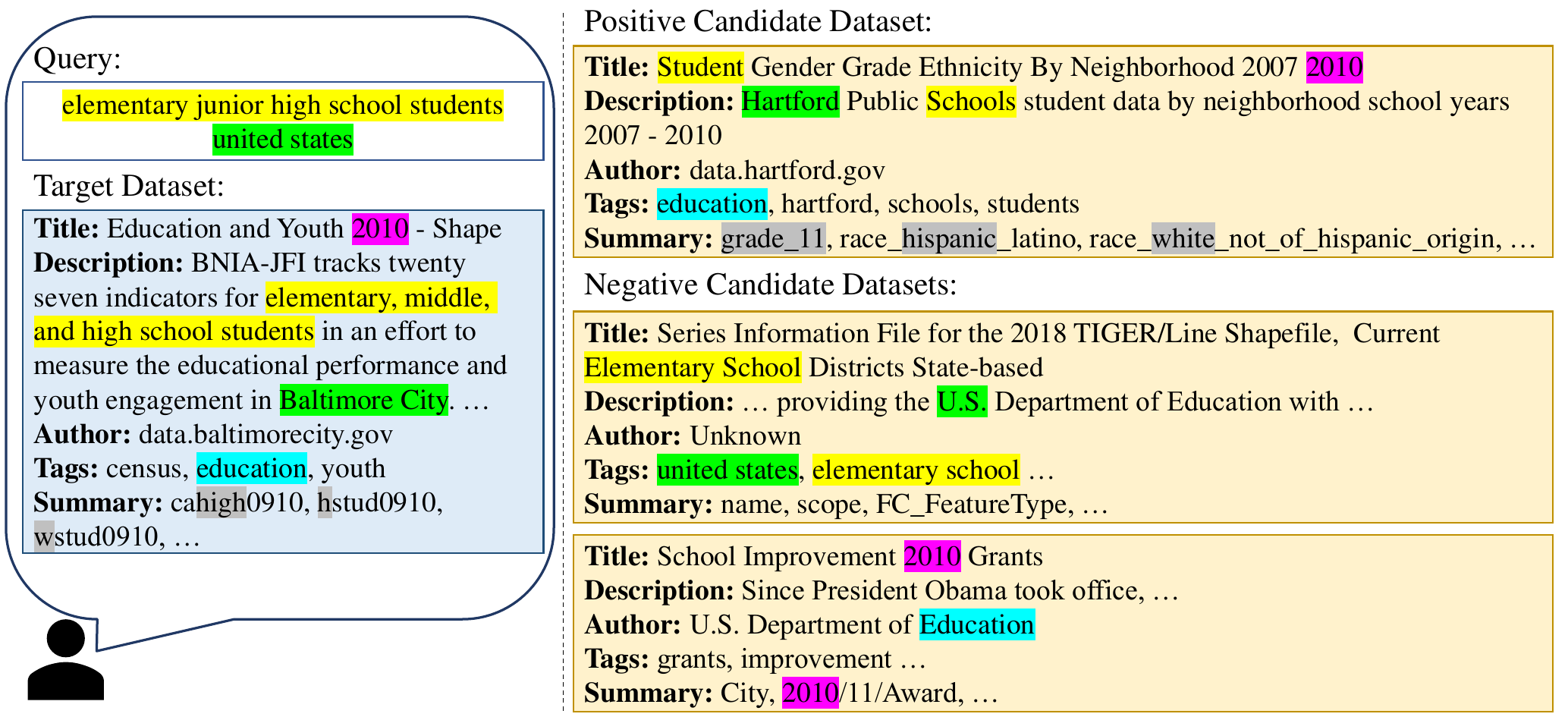}
\caption{An example of ExDSE. Left: The user submits a query and a target dataset. Right: The system aims to return positive candidate datasets that exhibit both relevance to the query (highlighted in yellow and green) and similarity to the target dataset (highlighted in magenta, cyan, and gray), as opposed to negative ones that are only relevant to the query or only similar to the target dataset. These judgments are supported by specific metadata and content fields.}
\label{fig:discovery-example}
\end{figure}

To address these limitations, Figure~\ref{fig:discovery-example} anticipates a combined scenario. A user poses a query ``elementary junior high school students united states'' while providing a target dataset ``Education and Youth 2010'' to imply specific preferences (e.g., student indicators in 2010). A positive candidate dataset aligns with both inputs,
while negative ones may match either input, but not both.
Moreover, the specific fields driving these relevance and similarity judgments are pinpointed.

\textbf{Task Definition.} We formulate the above scenario as a generalization of established paradigms, termed \emph{Dataset Search with Examples} (\textbf{DSE}). Given a keyword query $q$ and a set of target datasets $D_t$, the task is to retrieve a list of candidate datasets $D_c$ that are most relevant to $q$ and, simultaneously, similar to the datasets in $D_t$. As an extension, \emph{Explainable DSE} (\textbf{ExDSE}) requires identifying, for each candidate dataset $d \in D_c$, a subset of its metadata or content fields $F_d$ that indicate $d$'s relevance to $q$ or similarity to $D_t$.

\textbf{Contributions.} Progress in DSE research is hindered by the lack of test collections. Current benchmarks~\cite{ntcir,ACORDAR1,ACORDAR2,DBLP:conf/ecir/KolyadaPS25} are designed for keyword-only input and lack mechanisms to incorporate target datasets. Moreover, they provide dataset-level but not field-level annotations required for evaluating ExDSE, and suffer from data scarcity due to high annotation costs. To fill the gap, we present \textbf{DSEBench}, the first test collection for evaluating DSE and ExDSE:
\begin{itemize}
    \item It provides high-quality human annotations for query relevance and target similarity to enable DSE evaluation, fine-grained field-level annotations for ExDSE evaluation, and synthetic annotations for training.
    \item We adapt a variety of retrieval, reranking, and explanation methods and evaluate them on DSEBench to establish a performance baseline.
\end{itemize}
\section{Related Work}
\label{sec:related-work}

\subsection{Dataset Search Benchmarks}

Dataset search research broadly follows three paradigms.

Relevance-based approaches retrieve datasets based on textual relevance to a query, including general-purpose systems like Google Dataset Search~\cite{GoogleDatasetSearch} and specialized systems for RDF datasets~\cite{DBLP:conf/semweb/ChenHZLLSC23,dunks,CKGSE}. The evaluation is based on established test collections including NTCIR~\cite{ntcir}, ACORDAR~\cite{ACORDAR1,ACORDAR2}, and WTR~\cite{wtr}.

Similarity-based discovery, particularly popular in the database community, aims to find datasets similar to an input dataset~\cite{DBLP:journals/csur/PatonCW24}. Test collections like LakeBench~\cite{lakebench} provide ground truth for relational data tasks such as finding joinable or unionable tables~\cite{TUS,D3L,str}, not applicable to generalized dataset search where metadata plays a key role and data formats vary.

Combined approaches such as Aurum~\cite{aurum} integrate keyword search with similarity-based navigation. Despite the existence of such systems, there is no public test collection to systematically evaluate this composite task. \emph{Our DSEBench fills this critical gap and provides the first performance baseline for DSE.}

\subsection{Explainable Dataset Search}

Existing explainability research in this domain focuses primarily on the generation of data snippets to summarize dataset content~\cite{CDS,miu_DS}. Although effective for content understanding, ExDSE further requires identifying the specific metadata fields that indicate query relevance and target similarity. This can be framed as a feature attribution problem. Following the taxonomy of explainable IR~\cite{DBLP:journals/corr/abs-2211-02405,DBLP:conf/ecir/AnandSV25}, post hoc methods such as LIME~\cite{lime} and SHAP~\cite{shap} are applicable here to quantify field contributions. In this regard, \emph{DSEBench is among the first to provide field-level ground truth to evaluate ExDSE.}



\subsection{Relevance Feedback}

Relevance Feedback (RF) refines queries using user judgments on retrieved documents~\cite{DBLP:journals/jasis/SaltonB90,DBLP:conf/cikm/YuXC21,rocchio1971relevance}.
DSE is conceptually similar to RF in that both leverage user-provided examples to guide the search process. Despite this high-level similarity, \emph{there are fundamental differences between DSE and RF.} First, DSE processes the query and target datasets as a single initial input, unlike the iterative interaction in RF. Second, the target datasets in DSE define a desired semantic profile (e.g., specific provenance or context) that complements topicality. To capture such relationships, directly adapting RF methods to DSE is suboptimal, as revealed by our evaluation, reflecting the unique challenges of DSE.
\section{Construction of DSEBench}
\label{sec:test-collection}

We constructed DSEBench by extending the English subset of NTCIR~\cite{ntcir}, a popular test collection for keyword-based dataset search. The original collection offers 46,615~datasets, 92,930~data files, and associated relevance judgments.

\subsection{Dataset Preprocessing}

The NTCIR datasets offer semi-structured metadata and text-rich data files. The latter contains a wealth of information that is not covered by metadata and is crucial for effective search~\cite{DBLP:conf/cikm/ChenWCKQ19}, necessitating their inclusion in our test collection.

\textbf{Dataset Fields.}
We define a dataset by five fields: four standard \emph{metadata fields} provided by the publisher (\texttt{title}, \texttt{description}, \texttt{tags}, \texttt{author}) and a \emph{content field} (\texttt{summary}) extracted from the data files to compactly describe their content. When we refer to a dataset in this paper, we indicate its five fields.

\begin{wraptable}{r}{4.5cm}
\vspace{-1em}
\caption{Format Distribution of Data Files}
\centering
\small
\label{tab:format}
\begin{tabular}{lrr}
\toprule
Format    & \#Files & \%       \\
\midrule
PDF       & 47,413       & 51.02\%  \\
TXT       & 12,531       & 13.48\%  \\
HTML      & 9,298        & 10.01\%  \\
CSV       & 7,577        & 8.15\%   \\
XML       & 4,549        & 4.90\%   \\
JSON      & 1,940        & 2.09\%   \\
RDF       & 1,484        & 1.60\%   \\
XLSX, XLS & 1,466        & 1.58\%   \\
DOCX, DOC & 453          & 0.49\%   \\
Others    & 6,219        & 6.69\%   \\
\bottomrule
\end{tabular}
\end{wraptable}

\textbf{Content Summary.}
The \texttt{summary} field is essential because raw data files are challenging to process directly due to their diverse formats and large sizes. As shown in Table~\ref{tab:format}, file formats range from unstructured PDF to structured CSV and RDF. Direct use of data files in search is hindered by readability issues (e.g., machine-readable RDF) and excessive length. To address this, we generate content summaries for quick comprehension, ensuring that the content is usable for retrieval and readable for human annotators. Their effectiveness in improving retrieval accuracy has been demonstrated~\cite{CDS,miu_DS}.

The data formats offered by NTCIR are not always accurate. We detected the real format of each data file using \texttt{python-magic}. \emph{Format-specific content summaries} are defined below.
\begin{itemize}
    \item Tabular Files (CSV, XLSX, XLS): We extracted all \emph{table headers} to provide a concise description of the data structure and attributes.
    \item Textual Files (PDF, TXT, HTML, DOCX): We generated an \emph{abstract} to distill key ideas and topics from the text.
    \item Structured Files (JSON, XML, RDF): We extracted all \emph{keys}, \emph{tags}, or \emph{predicates} to compactly represent the data organization and schema.
    \item Others (HDF, etc.): Their summaries were left empty.
\end{itemize}

\textbf{Summary Generation.}
For tabular files, we applied heuristic rules to extract headers from CSV, achieving an accuracy of~92\% on a random sample of 50~files. For complex XLSX/XLS structures, we followed~\cite{TableHeaderDetection} to extract headers based on inherent features, achieving an accuracy of~91\% in 100~samples.

For textual files, we performed optical character recognition on PDF and removed HTML tags. We then used \texttt{bart-large-cnn}~\cite{bart} to generate abstracts, truncating the input text to 1,024~tokens and generating up to 300~tokens.

For structured files, we extracted keys, tags, and predicates from JSON, XML, and RDF using \texttt{json}, \texttt{xml}, and \texttt{rdflib}, respectively.

\subsection{Test and Training Cases}
\label{subsec:queries}

Recall that the input of DSE consists of a query~$q$ and a set of target datasets~$D_t$. We adapted NTCIR's query-dataset pairs to DSE as follows.

\textbf{Test and Training Split.}
We adapted the 141~query-dataset pairs annotated as highly relevant in NTCIR to form our \emph{test cases} where the query and the dataset became~$q$ and a singleton set~$D_t$ of
target datasets, respectively. For quality considerations, the 1,434~query-dataset pairs annotated as partially relevant in NTCIR were similarly adapted to our \emph{training cases}. To avoid train-test overlap, we filtered out any pairs that involved queries or datasets present in the test cases, resulting in 672~training cases.

\textbf{Synthetic Training Cases.}
Since 672~training cases would be insufficient for fine-tuning large models, we synthesized a larger number of training cases from 45,472~datasets in NTCIR that are not annotated relevant to any query. Specifically, we fine-tuned a \texttt{T5-base} model~\cite{t5base} to generate a query~$q$ from the title and description of each dataset, so that the input dataset served as~$D_t$. The model was fine-tuned on the $141+1,434$~pairs mentioned above with an 80/20 train/valid split for 20~epochs. We selected an optimal learning rate of $1e-5$ from $\{1e-6, 5e-6, 1e-5\}$ and a batch size of~8 from $\{8, 16\}$ based on the  average score of ROUGE-1, ROUGE-2, and ROUGE-L.

To improve the quality of synthetic queries, we applied rigorous filtering to remove trivial queries that duplicated the title of the dataset, verbose queries with excessive repetition of words, and vague queries that were too broad and lacked concrete search intent. We also ensured that the generated queries did not overlap with the test cases. This yielded 3,810~distinct queries generated from 5,065~datasets, bringing the total number of training cases from~672 to~5,737.

\subsection{Pooling}
\label{subsec:pooling}

It would be impractical to manually judge the relevance and similarity of 46,615 datasets to 141~test and 5,737~training queries. Following common practices in IR evaluation, we used the pooling method to reduce the number of judgments.

\textbf{Retrievers.}
We adapted four popular retrievers, including two lexical models (BM25, TF-IDF) and two dense models (\texttt{BGE-large-en-v1.5}~\cite{bge_embedding}, \texttt{GTE-large}~\cite{gte_embedding}).

\textbf{Implementation.}
We concatenated all fields of a dataset to construct a pseudo-document. Given an input $(q, D_t)$, since DSE requires finding datasets relevant to both~$q$ and~$D_t$, we concatenated~$q$ with the pseudo-document of each target dataset $d_t \in D_t$ into an expanded query to retrieve candidate datasets.\footnote{To prevent short but important fields from being overlooked, we followed~\cite{query2doc,QueryExpansion} to repeat the query, \texttt{tags}, and \texttt{author} fields 100~times during concatenation.}

\textbf{Results.}
We collected 20~candidate datasets with the highest rank from each retriever, resulting in 7,415~$(q, D_t, d_c)$ triples for test cases and 337,976~triples for training cases, where $d_c$~indicates a candidate dataset to be annotated. To balance between quality and cost of annotation, we invited human experts to annotate test cases and used a large language model (LLM) for training cases.

\subsection{Human Annotation}

We recruited 24~graduate or undergraduate computer science students with experience in dataset search to annotate 7,415~$(q, D_t, d_c)$ triples in test cases.

\textbf{Dataset-Level Judgments.}
Given a triple $(q, D_t, d_c)$, following the task definition of DSE, an annotator judged how much the candidate dataset~$d_c$ is \emph{relevant to the query}~$q$, and how much $d_c$~is \emph{similar to the target datasets} in~$D_t$. In accordance with NTCIR, the judgments were scored using a graded scale:
\begin{itemize}
    \item 0: irrelevant or dissimilar,
    \item 1: partially relevant or partially similar,
    \item 2: highly relevant or highly similar.
\end{itemize}

\textbf{Field-Level Judgments.}
For each partially or highly relevant/similar dataset, the annotator was further required to identify in the candidate dataset \emph{which fields indicate its relevance/similarity}, to be used in the evaluation of ExDSE.

\begin{figure}[t]
\centering
\includegraphics[width=\linewidth]{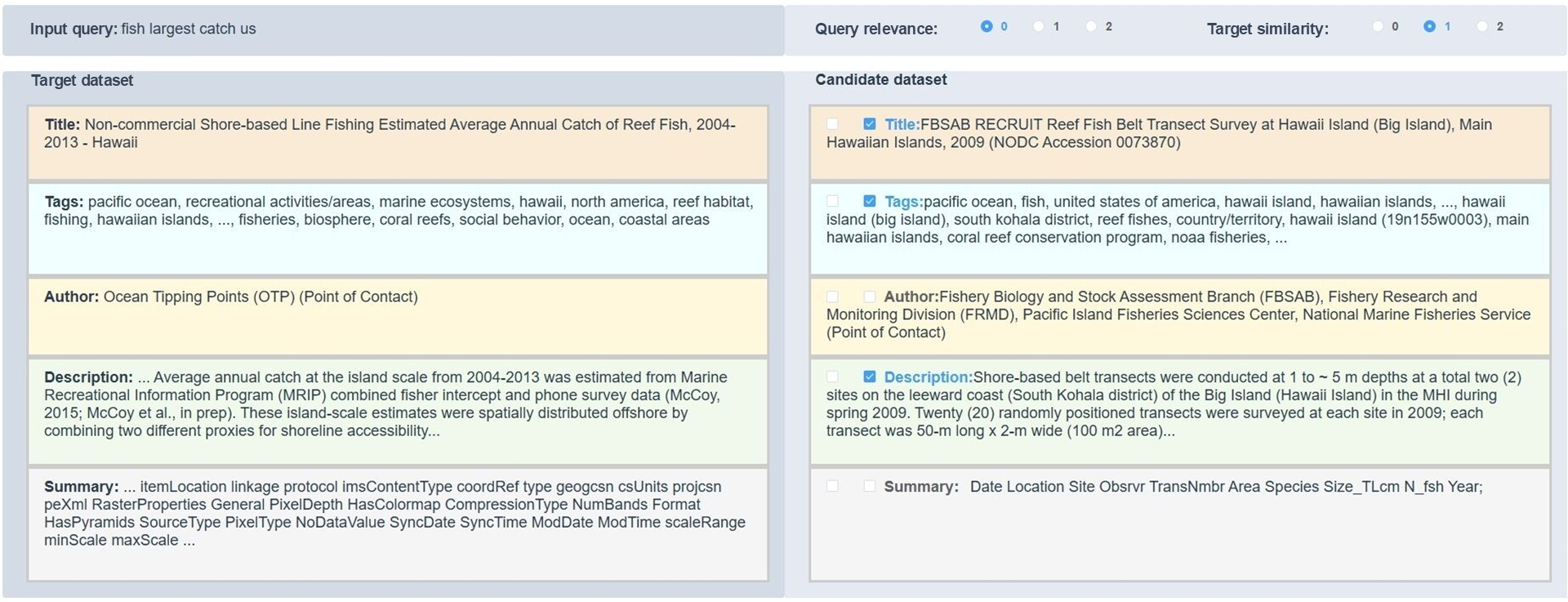}
\caption{A screenshot of our tool for human annotation. Left: An input of DSE, including a query and a target dataset. Right: A candidate dataset with annotation options. In this example, the candidate dateset is annotated partially similar to the target dataset. Accordingly, the second column of checkboxes are enabled, and three fields of the candidate dataset are annotated indicators of its similarity. However, the candidate dateset is annotated irrelevant to the query, so the first column of checkboxes representing indicators of relevance are disabled.}
\label{fig:annotation-example}
\end{figure}

\textbf{Annotation Tool.} Annotation was performed using a Web tool illustrated in Figure~\ref{fig:annotation-example}. To avoid position bias, the fields were displayed in random order.

\textbf{Task Assignment.}
To obtain high-quality annotations, we assigned each of the 7,415~$(q, D_t, d_c)$ triples from test cases to \emph{two independent annotators}. When they provided the same score, it was accepted as the final score. Otherwise, we assigned the triple to a \emph{third annotator}, and the final score was determined on the basis of a majority vote, that is, by the third annotator. However, when three scores differed from each other, we took their mean value (i.e.,~1) as the final score. Field-level disagreements were resolved analogously.

\begin{table}[t]
\caption{Distribution of Dataset-Level Judgments}
\label{tab:relevance-distribution}
\centering
\small
\begin{tabular}{lrrccc}
\toprule
                         & \#Cases          & \#Triples            & Judgment            & Query Relevance & Target Similarity \\
\midrule
\multirow{3}{*}{Test Cases} & \multirow{3}{*}{141}   & \multirow{3}{*}{7,415} & 0  & 5,056 (68.19\%)  & 3,891 (52.47\%)  \\ \cmidrule(lr){4-6}
                                  &                        &           & 1 & 1,622 (21.87\%)  & 2,474 (33.36\%)  \\ \cmidrule(lr){4-6}
                                  &                        &           & 2    & 737 (9.94\%)  & 1,050 (14.16\%)  \\
\midrule
\multirow{3}{*}{Training Cases} & \multirow{3}{*}{5,699} & \multirow{3}{*}{122,585} & 0         & 67,541 (55.10\%)  & 58,592 (47.80\%)  \\ \cmidrule(lr){4-6}
                                  &                        &            & 1 & 41,290 (33.68\%)  & 52,590 (42.90\%)  \\ \cmidrule(lr){4-6}
                                  &                        &            & 2    & 13,754 (11.22\%)  & 11,403 (9.30\%)  \\
\bottomrule
\end{tabular}
\end{table}

\begin{table}[t]
\caption{Distribution of Field-Level Judgments}
\label{tab:field-level-judgments-distribution}
\centering
\small
\begin{tabular}{lrrcrc}
\toprule
                                & \multicolumn{3}{c}{Query Relevance} & \multicolumn{2}{c}{Target Similarity} \\
\cmidrule(lr){2-4}  \cmidrule(lr){5-6}
 & \#Triples              & \multicolumn{2}{r}{Indicator Field} & \#Triples              & Indicator Field  \\
\midrule
\multirow{5}{*}{Test Cases}
& \multirow{5}{*}{2,359}  & title:         & 2,086 (88.43\%)    & \multirow{5}{*}{3,524}  & 2,941 (83.46\%)  \\
\cmidrule(lr){3-4} \cmidrule(lr){6-6}
                                &                         & description:   & 2,132 (90.38\%)    &                         & 2,818 (79.97\%)  \\
\cmidrule(lr){3-4} \cmidrule(lr){6-6}
                               &                         & tags:          & 1,135 (48.11\%)    &                         & 2,206 (62.60\%)  \\
\cmidrule(lr){3-4} \cmidrule(lr){6-6}
                                &                         & author:        & 153 (6.49\%)       &                         & 1,916 (54.37\%)  \\
\cmidrule(lr){3-4} \cmidrule(lr){6-6}
                               &                         & summary:       & 373 (15.81\%)      &                         & 882 (25.03\%)    \\
\midrule
\multirow{5}{*}{Training  Cases}
& \multirow{5}{*}{55,044} & title:         & 53,233 (96.71\%)   & \multirow{5}{*}{63,993} & 18,273 (28.55\%) \\
\cmidrule(lr){3-4} \cmidrule(lr){6-6}
                                &                         & description:   & 41,666 (75.70\%)   &                         & 18,791 (29.36\%) \\
\cmidrule(lr){3-4} \cmidrule(lr){6-6}
                                &                         & tags:          & 35,141 (63.84\%)   &                         & 42,428 (66.30\%) \\
\cmidrule(lr){3-4} \cmidrule(lr){6-6}
                                &                         & author:        & 13,540 (24.60\%)   &                         & 25,504 (39.85\%) \\
\cmidrule(lr){3-4} \cmidrule(lr){6-6}
                                &                         & summary:       & 8,950 (16.26\%)    &                         & 23,902 (37.35\%) \\
\bottomrule
\end{tabular}
\end{table}

\textbf{Results.}
Table~\ref{tab:relevance-distribution} presents the distribution of the judgments made at the dataset level. As shown in the upper half of the table, more than two thirds~(68.19\%) of the candidate datasets are irrelevant to the query and more than half~(52.47\%) are dissimilar to the target datasets. The highly relevant and similar datasets account for~9.94\% and~14.16\%, respectively.

Table~\ref{tab:field-level-judgments-distribution} presents the distribution of the judgments made at the field level. As shown in the upper half of the table, two popular indicator fields are \texttt{title} and \texttt{description}. Although \texttt{author} is rarely selected as an indicator field for query relevance~(6.49\%), it often indicates target similarity~(54.37\%). Another fairly useful field is \texttt{summary} (15.81\%--25.03\%).

\begin{figure}[t]
\centering
    \begin{subfigure}[b]{0.43\columnwidth}
        \centering
        \includegraphics[width=\textwidth]{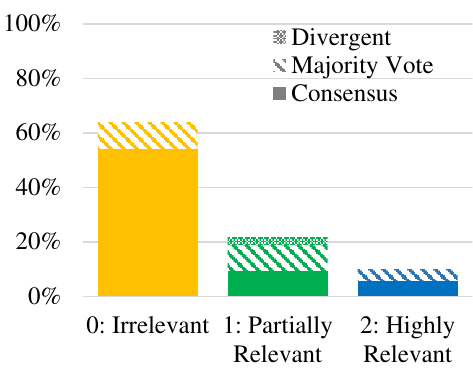}
        \caption{Query Relevance}
        \label{fig:rel-score-distribution-query}
    \end{subfigure}
    \hfill
    \begin{subfigure}[b]{0.43\columnwidth}
        \centering
        \includegraphics[width=\textwidth]{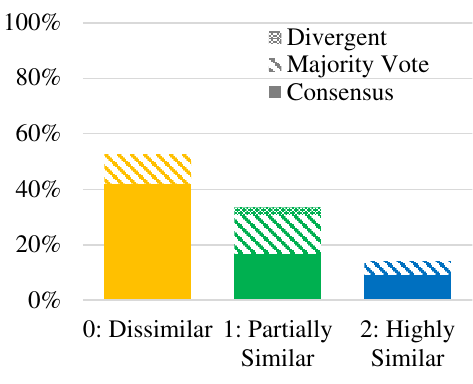}
        \caption{Target Similarity}
        \label{fig:sim-score-distribution-target}
    \end{subfigure}
\caption{Inter-annotator agreements on dataset-level judgments for the test cases.}
\label{fig:distribution-of-human-annotation}
\end{figure}

\begin{figure}[t]
\centering
    \begin{subfigure}[b]{0.43\columnwidth}
        \centering
        \includegraphics[width=\textwidth]{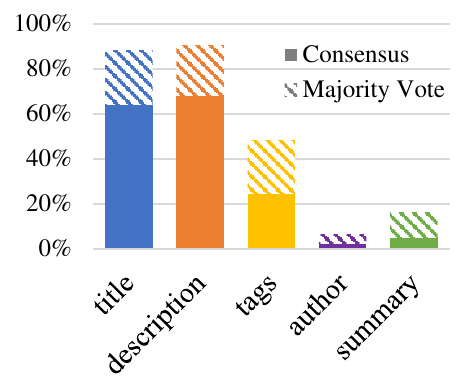}
        \caption{Query Relevance}
        \label{fig:annotated-field-percentage-query}
    \end{subfigure}
    \hfill
    \begin{subfigure}[b]{0.43\columnwidth}
        \centering
        \includegraphics[width=\textwidth]{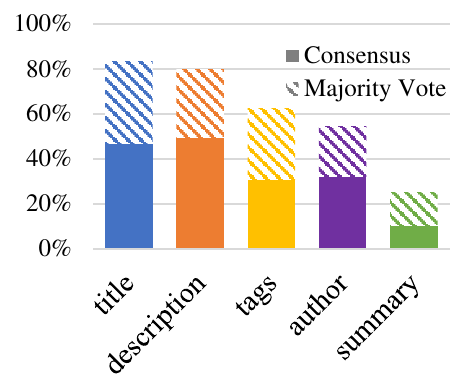}
        \caption{Target Similarity}
        \label{fig:annotated-field-percentage-target}
    \end{subfigure}
\caption{Inter-annotator agreements on field-level judgments for the test cases.}
\label{fig:distribution-of-human-annotation-field}
\end{figure}

\textbf{Inter-Annotator Agreement.}
Figure~\ref{fig:distribution-of-human-annotation} shows the distribution of inter-annotator agreements on the dataset-level judgments. Two annotators reach a consensus on the query relevance and target similarity for more than~69\% and~68\% of the triples, respectively. Only for less than~3\% of the triples, a third annotator is introduced but the three judgments are still divergent, so their mean value must be taken. We calculated Krippendorff's~$\alpha$, a statistical measure of the agreement achieved. The results $\alpha=0.51$ for query relevance and $\alpha=0.52$ for target similarity represent \emph{a moderate degree of agreement between the annotators}, higher than $\alpha=0.44$ reported on NTCIR~\cite{ntcir}.

Figure~\ref{fig:distribution-of-human-annotation-field} shows the distribution of inter-annotator agreements on the field-level judgments. Two annotators reach a consensus for more than~65\% and~55\% of the judgments of indicator field for query relevance and target similarity, respectively. Their disagreements are resolved by introducing a third annotator.

\subsection{LLM Annotation}
\label{sec:llmann}

We prompted a popular and cost-effective LLM, GLM-3-Turbo~\cite{glm2024chatglm}, to annotate 337,976~$(q, D_t, d_c)$ triples in training cases. In accordance with human annotation, we collected dataset-level and field-level judgments from LLM outputs.

\textbf{Quality Control.}
We found that LLM provided many superficially inaccurate dataset-level judgments, so we applied heuristic filtering. For query relevance, we removed the cases where the occurrence of query keywords in candidate datasets was either implausibly low despite being labeled relevant or implausibly high despite being labeled irrelevant. For target similarity, we removed the cases where the overlap between the candidate dataset and the target dataset was inconsistent with the label, e.g.,~for a dissimilar pair, the string similarity of their fields was too high or they shared an excessive number of tags.

Although these heuristics might also mistakenly filter out high-quality judgments, we checked the results and, in general, observed \emph{substantially improved quality}. The accuracy for query relevance and target similarity judgments increased from~73\% and~69\%, respectively, to more than~85\% based on a manual check of 3,378 triples sampled. Given this satisfactory quality and large size of LLM annotations, they form useful training cases for weak supervision.

We did not filter field-level judgments because our investigation revealed their relatively high accuracy but complex error patterns that are hard to detect.

\textbf{Results.}
Finally, we kept 122,585~triples with complete annotations (i.e., no judgment discarded) from 5,699~training cases. The lower half of Tables~\ref{tab:relevance-distribution} and~\ref{tab:field-level-judgments-distribution} present their distributions of dataset-level and field-level judgments, respectively, generally resembling the distributions in test cases.
\section{Evaluation of DSE}
\label{sec:experiments-on-dataset-discovery}

A typical IR pipeline consists of \emph{first-stage retrieval} and \emph{reranking}. Following this convention, we used DSEBench to separately evaluate retrieval and reranking models. All experiments were conducted on an Intel Xeon Gold 5222 CPU with four NVIDIA GeForce RTX 3090 GPUs.
Our code is on GitHub.\footnote{\url{https://github.com/nju-websoft/DSEBench}}

\subsection{Training, Validation, and Test Sets}
\label{subsec:data-preprocessing}

DSEBench provides 141~test cases and 5,699~training cases. To ensure comparability, we provide two official splits:
\begin{itemize}
    \item \textbf{Five-Fold Split:} We randomly partitioned 141~test cases into five nearly equal-sized folds for cross-validation (3/1/1 for training/validation/test). Performance should be averaged over five runs.
    \item \textbf{Annotator Split:} We divided the 5,699~training cases into~80\% for training and~20\% for validation, using the 141~test cases strictly for testing.
\end{itemize}

\subsection{Evaluation Metrics}

Standard IR evaluation metrics were used on the test set: MAP@5, NDCG@5, Recall@5 (R@5), MAP@10, NDCG@10, and Recall@10 (R@10), where NDCG@10 was also used for the selection of hyperparameters on the validation set. All metrics were calculated using \texttt{pytrec-eval-terrier}.

These metrics rely on gold-standard labels. Recall that each triple $(q, D_t, d_c)$ annotated in DSEBench is associated with two dataset-level judgments: the relevance of the candidate dataset~$d_c$ to the query~$q$ and the similarity of~$d_c$ to the target datasets in~$D_t$, both in the range of $\{0,1,2\}$. Following the task definition of DSE, since query relevance and target similarity are indispensable, their product was taken as the gold-standard graded label of $(q, D_t, d_c)$ in the range of $\{0,1,2,4\}$ to ensure that $d_c$~receives a high overall score only if it is highly relevant to~$q$ \emph{and} highly similar to~$D_t$; a low score in either aspect would significantly penalize the overall score. Note that query relevance and target similarity can also be combined in other reasonable ways, such as their harmonic mean.

\subsection{Benchmarking Retrieval Models}
\label{subsec:retrieval}

\textbf{Retrievers.} 
We adapted and evaluated seven representative retrieval models:
\begin{itemize}
    \item unsupervised models, including lexical models \textbf{BM25} and \textbf{TF-IDF}, and dense models \textbf{BGE} (\texttt{bge-large-en-v1.5})~\cite{bge_embedding} and \textbf{GTE} (\texttt{gte-large})~\cite{gte_embedding};
    \item supervised models, including \textbf{DPR}~\cite{dpr} (bi-encoder), \textbf{ColBERTv2}~\cite{colbertv2} (late interaction), and \textbf{coCondenser}~\cite{cocondenser} (specialized pre-training).
\end{itemize}

We also evaluated a classic RF method by adapting it to the DSE task:
\begin{itemize}
    \item Rocchio~\cite{rocchio1971relevance} uses target datasets as positive feedback (\textbf{Rocchio-P}) or also infers negative feedback from non-target datasets (\textbf{Rocchio-PN})~\cite{DBLP:conf/cikm/OosterhuisR18}.
\end{itemize}

\textbf{Implementation Details.}
Following the implementation in Section~\ref{subsec:pooling}, we concatenated all fields to construct pseudo-documents as dataset representations. The query was expanded with the pseudo-documents of target datasets.

The supervised models were evaluated in three configurations: \textbf{not fine-tuned}, \textbf{fine-tuned with the five-fold split}, and \textbf{fine-tuned with the annotator split}, as defined in Section~\ref{subsec:data-preprocessing}.
Since fine-tuning requires binary labels, we mapped the gold-standard graded scores in the range of $\{0,1,2,4\}$ to binary labels: positive or zero. For BERT-based models, we set their maximum length of the input sequence to 512~tokens. During fine-tuning, we performed grid search to select learning rate from $\{1e-5, 5e-6\}$ and batch size from $\{8, 16\}$. As a result, with the five-fold split, we selected a learning rate of~$1e-5$ and a batch size of~8 for DPR, and a learning rate of~$1e-5$ and a batch size of~16 for ColBERTv2 and coCondenser. With the annotator split, we selected a learning rate of~$1e-5$ and a batch size of~16 for DPR, and a learning rate of~$1e-5$ and a batch size of~8 for ColBERTv2 and coCondenser. For RF methods, the weights to combine the vector representations of the original query, positive feedback, and negative feedback to refine the query vector were~1.0, 0.75, and~0.15, respectively.

\begin{table}[t]
\caption{Evaluation Results of Retrieval Models}
\label{tab:retrieval-results}
\small
\centering
\begin{tabular}{lcccccc}
\toprule
              & MAP@5                             & @10                           & NDCG@5                       
     & @10                          & R@5  & @10 \\
\midrule
\multicolumn{7}{l}{\textit{Unsupervised Models}}                    \\
BM25                                           
 & 0.0982                            & 0.1739                           & 0.3059                           & 0.3416              
              & 0.1705                           & 0.2769                           \\
TF-IDF                               
           & 0.0921                            & 0.1615                           & 0.2971                           & 0.3227    
                        & 0.1572                           & 0.2576                           \\
BGE                     
                        & 0.1045                            & 0.1834                           & 0.3233                   
         & 0.3598                           & 0.1756                           & 0.2887                           \\
GTE        
                                     & 0.1104                            & 0.1921                           & 0.3267      
                      & 0.3649                           & 0.1820                            & 0.2983                     
       \\
\midrule
\multicolumn{7}{l}{\textit{Supervised Models (not fine-tuned)}}  \\
DPR                                            & 0.1072                            & 0.1757               
             & 0.3248                           & 0.3472                           & 0.1706                           & 0.2695   
                         \\
ColBERTv2                                      & 0.1072                            & 0.1799        
                    & 0.3206                           & 0.3510                            & 0.1687                       
     & 0.2720                            \\
coCondenser                                    & 0.1029                            & 0.1671 
                           & 0.3142                           & 0.3326                           & 0.1597                 
           & 0.2525                           \\
\midrule
\multicolumn{7}{l}{\textit{Supervised Models (fine-tuned with the five-fold split)}}    \\
DPR                                            & 0.1204       
                      & 0.1929                           & 0.3574                           & 0.3769                      
      & 0.1833                           & 0.2967                           \\
ColBERTv2                                      & 
 0.1247                            & 0.2052                           & 0.3449                           & 0.3779               
             & 0.1909                           & 0.3066                           \\
coCondenser & \textbf{0.1387} & \textbf{0.2286} & \textbf{0.3784} & \textbf{0.4147} & \textbf{0.2110} & \textbf{0.3401}          \\
\midrule
\multicolumn{7}{l}{\textit{Supervised Models (fine-tuned with the annotator split)}} \\
DPR   
                                          & 0.1194                            & 0.1936                           & 0.3551 
                           & 0.3757                           & 0.1851                           & 0.2914                 
           \\
ColBERTv2                                      & 0.1109                            & 0.1925                      
      & 0.3360                            & 0.3760                            & 0.1800                             & 0.3033      
                      \\
coCondenser                                    & \underline{0.1286}             & \underline{0.2238}               & \underline{0.3656}            
      & \underline{0.4136}              & \underline{0.1979}                 & \underline{0.3367} \\
\midrule
\multicolumn{7}{l}{\textit{RF Methods}} \\
Rocchio-P    & 0.1033      & 0.1782       & 0.3139       & 0.3452        & 0.1713    & 0.2837    \\
Rocchio-PN     & 0.1023      
 & 0.1773       & 0.3125       & 0.3441        & 0.1696    & 0.2828     \\
\bottomrule
\end{tabular}
\end{table}

\textbf{Evaluation Results.}
We evaluated the highest-ranked datasets retrieved by each model from the full corpus of 46,615~datasets.
Table~\ref{tab:retrieval-results} presents the results. Among unsupervised models, dense models (BGE and GTE) outperform lexical models (BM25 and TF-IDF). For supervised models, fine-tuning yields substantial gains. The performance difference between the five-fold and annotator splits is minimal, suggesting that the \emph{large-scale LLM annotations in DSEBench are almost as effective as human annotations for training}. RF methods slightly improve on the vanilla TF-IDF, but remain inferior to other approaches, highlighting that \emph{adapting RF is insufficient for addressing the complex DSE task}.

\subsection{Benchmarking Reranking Models}
\label{subsec:benchmarking-reranking-models}

\textbf{Rerankers.} 
We adapted and evaluated six representative reranking models:
\begin{itemize}
    \item text-based models, including the state of the art on MTEB's reranking leaderboard such as \textbf{SFR}\texttt{-Embedding-Mistral}~\cite{SFRAIResearch2024} and \textbf{Stella}\texttt{-en-1.5B-v5}~\cite{Stella}, and \textbf{BGE-reranker}\texttt{-v2-minicpm-layerwise}~\cite{bge_m3}, a supervised cross-encoder model specifically designed for reranking tasks;
    \item structure-based models, including \textbf{HINormer}~\cite{HINormer}, a popular heterogeneous Graph Neural Network (GNN), and the state-of-the-art \textbf{HHGT}~\cite{HHGT}, both employing a graph structure representing relationships between datasets~\cite{google_relationships};
    \item \textbf{LLM}, specifically \texttt{GLM-4-Plus}~\cite{glm2024chatglm}, which supports a long context length.
\end{itemize}

\textbf{Implementation Details.}
For text-based models, the textual representations of the datasets and queries were constructed in the same way as in Section~\ref{subsec:retrieval}. We did not fine-tune SFR and Stella. BGE-reranker was evaluated in the same way as the supervised models in Section~\ref{subsec:retrieval}: with the five-fold split, we selected a learning rate of~$1e-5$ and a batch size of~16; with the annotator split, we selected a learning rate of~$1e-5$ and a batch size of~8.

For structure-based GNN models, we constructed a graph in which the nodes represent datasets and the edges represent relationships between datasets~\cite{google_relationships}. GNN was performed on this graph to enhance the textual representations of the dataset nodes. To predict the reranking score of a dataset, we fed into a multi-layer perceptron its enhanced node representation and the mean representation of top-20 nodes whose initial textual representations best matched the query. We trained the models with a learning rate of $1e-4$, setting the batch size to~64 with the five-fold split and 256 with the annotator split. Regarding the model architecture, we set $\texttt{num-gnns}=1$ and $\texttt{num-layers}=2$ for HINormer, and set $L\_hop=2$, $L\_type=2$, and $hops=1$ for HHGT.

LLM was evaluated in four configurations: \textbf{zero-shot} and \textbf{one-shot} which directly rerank all datasets, \textbf{RankLLM} (\texttt{ListwiseRankLLM})~\cite{RankLLM} which performs listwise reranking, and \textbf{multi-layer}~\cite{multi_layer} which repeatedly (20~times) divides the datasets into five groups and reranks each group, and then reranks all datasets by their frequency of appearing in the top half of a reranked group.

\begin{table}[t]
\small
\caption{Evaluation Results of Reranking Models}
\label{tab:reranking-results}
\centering
\begin{tabular}{lcccccc}
\toprule
              & MAP@5                             & @10                           & NDCG@5                      
      & @10                          & R@5  & @10 \\
\midrule
\multicolumn{7}{l}{\textit{Supervised Models (not fine-tuned)}}                    \\
Stella                                 & 0.1180      
                               & 0.2106                           & 0.3509                           & 0.3981             
               & 0.1938                           & 0.3307                           \\
SFR                              
       & 0.1184                                    & 0.2090                           & 0.3488                           & 
 0.3920                           & 0.1940                           & 0.3225                           \\
BGE-reranker                 
           & 0.1170                                    & 0.2032                           & 0.3470                        
    & 0.3877                           & 0.1930                           & 0.3163                           \\
\midrule
\multicolumn{7}{l}{\textit{Supervised Models (fine-tuned with the five-fold split)}}       
              \\
BGE-reranker                           & 0.1178                                    & 0.2085                      
      & 0.3464                           & 0.3956                           & 0.1914                           & 0.3273          
                  \\
HINormer                           & 0.1009                                    & 0.1734                  
          & 0.3112                           & 0.3407                           & 0.1313                           & 0.2643      
                      \\
HHGT                           & 0.0874                                    & 0.1550              
              & 0.2781                           & 0.3090                           & 0.1139                           & 0.2333  
                          \\
\midrule
\multicolumn{7}{l}{\textit{Supervised Models (fine-tuned with the annotator split)}}                    \\
BGE-reranker                           & \underline{0.1249}                 & \underline{0.2146}   
               & \underline{0.3665}               & \underline{0.4099}             & \underline{0.1985}           & \underline{0.3347}          \\
HINormer                           & 0.0876     
                                & 0.1617                           & 0.2682                           & 0.3089            
                & 0.1142                           & 0.2339                           \\
HHGT                           & 0.0816 
                                    & 0.1449                           & 0.2658                           & 0.2977        
                    & 0.1215                           & 0.2342                           \\
\midrule
\multicolumn{7}{l}{\textit{LLMs}}                    \\
LLM (zero-shot)     
                          & 0.1144                                    & 0.1748                           & 0.3130         
                   & 0.3360                           & 0.1691                           & 0.2652                         
   \\
LLM (one-shot)                               & 0.1154                                    & 0.1776                           & 0.3058 
                           & 0.3327                           & 0.1691                           & 0.2729                 
           \\
LLM (RankLLM)               & 0.1191 & 0.2011 & 0.3506 & 0.3804 & 0.1735 & 0.2910 \\
LLM (multi-layer)               & \textbf{0.1468} & \textbf{0.2398} & \textbf{0.4071} & \textbf{0.4451} & \textbf{0.2093} & \textbf{0.3608} \\
\bottomrule
\end{tabular}
\end{table}

\textbf{Evaluation Results.}
We evaluated the highest-ranked datasets after reranking the union of the top-10 datasets retrieved by the models evaluated in Section~\ref{subsec:retrieval}. Table~\ref{tab:reranking-results} presents the results.
Among supervised models, BGE-reranker performs better after fine-tuning, particularly with the annotator split, \emph{reinforcing the value of the LLM annotations in DSEBench.}
The structure-based models (HINormer and HHGT) generally underperform compared to BGE-reranker, suggesting that while semantic relationships between datasets may be useful, textual representations remain the dominant signal for DSE.
For LLMs, the multi-layer setting achieves the highest scores across all metrics, likely thanks to the robustness of ensemble voting and reduced context length per inference.
\section{Evaluation of ExDSE}
\label{sec:explanation-experiments}

Building on DSE, ExDSE further requires identifying the indicator fields for query relevance and target similarity. We used the test cases in DSEBench to evaluate the explanation methods.

\textbf{Evaluation Metric.}
The F1 score was used to compare the set of indicator fields identified by an explanation method with the gold standard in DSEBench.

\textbf{Explainers.}
We adapted and evaluated four post hoc explainers:
\begin{itemize} 
    \item \textbf{Feature Ablation}, assessing the importance of a field based on the change in ranking scores after its removal;
    \item surrogate models, including two popular implementations \textbf{LIME}~\cite{lime} and \textbf{SHAP}~\cite{shap}, which train proxy models to estimate field contributions;
    \item \textbf{LLM}, specifically \texttt{GLM-4-Air}~\cite{glm2024chatglm}, directly generating indicator fields.
\end{itemize}

\textbf{Implementation Details.}
For Feature Ablation, we excluded one field at a time and calculated the ratio of the resulting ranking score to the original score. If the ratio for a field dropped below a threshold of~0.95, we outputted it as an indicator field. If no field met this threshold, we selected the field with the largest score drop as an indicator field.

LIME relies on a binary classifier as a proxy model to approximate the behavior of the retrieval model; we used SVM. Since a dataset has five fields, we set the number of perturbed samples used to train the classifier to a sufficiently large value of~50. Fields with a predicted negative score were outputted as indicator fields. For SHAP, we chose its partition explainer as the explainer algorithm. Fields with a positive Shapley value were outputted as indicator fields.

LLM was evaluated in two configurations: \textbf{zero-shot} and \textbf{few-shot} which included three verified examples selected from the training cases in DSEBench.

Since most of the explanation methods tested are coupled with the concrete retrieval model, we combined them with each of the retrieval models evaluated in Section~\ref{subsec:retrieval}, not including the RF method for its incompatibility.



\begin{table}[t]
\centering
\caption{Evaluation Results (F1) of Explanation Methods}
\label{tab:explanation-results}
\resizebox{\linewidth}{!}{
\begin{tabular}{lcccccccc}
\toprule
& \multicolumn{4}{c}{Query Relevance} & \multicolumn{4}{c}{Target Similarity} \\
\cmidrule(lr){2-5} \cmidrule(lr){6-9}
& \multirow{2}{*}{BM25} & \multirow{2}{*}{DPR} & \multirow{2}{*}{\parbox{1.5cm}{coCond.
\\ (f.-f.)}} & \multirow{2}{*}{\parbox{1.5cm}{coCond.\\ (ann.)}} & \multirow{2}{*}{BM25} & \multirow{2}{*}{DPR} & \multirow{2}{*}{\parbox{1.5cm}{coCond.
\\ (f.-f.)}} & \multirow{2}{*}{\parbox{1.5cm}{coCond.\\ (ann.)}} \\
&                       &                      &                                                  
   &                                                  &                       &                      &   
                                                  &                                                  \\
\midrule
Feature Ablation 
      & 0.4819                & 0.2806               & 0.4754                                             & 0.4193               
                             & 0.4750                & 0.3216               & 0.3675                                      
        & 0.3351                                           \\
LIME          & 0.6325                & \underline{0.6943}   & \underline{0.6960}                 
                 & \underline{0.7100}                               & 0.7562                & 0.7650               & \underline{0.7761}                  
                & \underline{0.7529}                               \\
SHAP          & \underline{0.6507}    & 0.6718               & 0.6761                     
                         & 0.6894                                           & \textbf{0.7835}       & \textbf{0.8201}      & \textbf{0.8007}                
                     & \textbf{0.7899}                                  \\
LLM (zero-shot) & 0.6455                & 0.6492               & 0.6351         
                                     & 0.6334                                           & 0.7315                & 0.7450  
              & 0.7264                                             & 0.7121                                        
    \\
LLM (few-shot)  & \textbf{0.7246}       & \textbf{0.7316}      & \textbf{0.7218}                                    & \textbf{0.7239}                                  & \underline{0.7647}    & \underline{0.7757} 
   & 0.7637                                             & 0.7508                                           \\
\bottomrule
\end{tabular}
}
\vspace{1ex}
\begin{minipage}{\linewidth}
\footnotesize
Note: DPR = DPR (not fine-tuned);
coCond. (f.-f.) = coCondenser (fine-tuned with the five-fold split); coCond. (ann.) = coCondenser (fine-tuned with the annotator split).
\end{minipage}
\end{table}

\textbf{Evaluation Results.}
Among the top-20~datasets retrieved by a retrieval model, we applied each explanation method to all datasets that have at least one annotated indicator field in DSEBench. Table~\ref{tab:explanation-results} presents the evaluation results averaged over all such datasets, obtained in a variety of retrieval model and configuration settings: unsupervised BM25, DPR not fine-tuned, coCondenser fine-tuned with the five-fold split, and coCondenser fine-tuned with the annotator split. The results with other retrieval models or configurations exhibit similar patterns, and thus are omitted.

Feature Ablation consistently performs worst, suggesting that simple field exclusion is ineffective for ExDSE.
Between surrogate models, SHAP outperforms LIME (and all others) in identifying indicator fields for target similarity, likely because of the robustness of its Shapley value theory compared to LIME's sampling mechanism.
LLM performs significantly better in the few-shot setting, exceeding other methods in identifying indicator fields for query relevance.

\begin{table}[t]
\caption{Field Preferences of Explanation Methods}
\label{tab:field-frequency}
\small
\centering
\begin{tabular}{ll}
\toprule
& Field Preferences by Frequency                                                                     \\
\midrule
Feature Ablation        & \texttt{description} \textgreater~ \texttt{title} \textgreater~ 
\texttt{summary} \textgreater~ \texttt{tags} \textgreater~ \texttt{author} \\
LIME            & \texttt{description} \textgreater~ \texttt{tags} \textgreater~ \texttt{title} \textgreater~ \texttt{summary} \textgreater~ \texttt{author} \\
SHAP            & \texttt{description} \textgreater~ \texttt{summary} \textgreater~ \texttt{title} \textgreater~ \texttt{tags} \textgreater~ \texttt{author} \\
LLM (zero-shot) & \texttt{description} \textgreater~ \texttt{tags} \textgreater~ \texttt{summary} \textgreater~ \texttt{title} \textgreater~ \texttt{author} \\
LLM (few-shot)  & \texttt{description} \textgreater~ \texttt{tags} \textgreater~ \texttt{title} \textgreater~ \texttt{summary} \textgreater~ \texttt{author} \\
\bottomrule
\end{tabular}
\end{table}

Table~\ref{tab:field-frequency} shows the general field preferences of each explanation method, where the fields are ordered by their frequency of being identified as an indicator field.
All methods prioritize \texttt{description}, likely due to its rich information.
In contrast, \texttt{author} is rarely selected.
In particular, the extracted content \texttt{summary} has a selection frequency comparable to key metadata such as \texttt{title} and \texttt{tags}, \emph{confirming the utility of content summaries in ExDSE.}
\section{Conclusions}
\label{sec:conclusions}

This paper investigates DSE, a composite IR task that generalizes two established dataset search paradigms: keyword-based search and similarity-based discovery. To advance research in this area, we introduce and release DSEBench, the first dedicated test collection for DSE, comprising high-quality human-annotated test cases alongside a large set of LLM-annotated training cases, supporting the evaluation of both unsupervised and supervised methods. The annotations encompass not only query relevance and target similarity of datasets in DSE, but also indicator fields that serve as explanations, thereby supporting ExDSE evaluation. Using DSEBench, we adapt a variety of existing retrieval, reranking, and explanation methods and evaluate their performance to establish comprehensive baselines for future comparative research.

\textbf{Impact.}
DSEBench has significant potential across diverse research communities, including the Semantic Web~\cite{DBLP:conf/semweb/ChenHZLLSC23,google_relationships,dunks}, databases~\cite{DBLP:journals/vldb/ChapmanSKKIKG20,DBLP:journals/csur/PatonCW24}, and IR~\cite{GoogleDatasetSearch,CDS,miu_DS}, where dataset search has been a long-standing focus. By supporting the dual tasks of DSE and ExDSE, DSEBench is positioned to serve as a reusable test collection to advance future research in retrieval and explanation techniques. This, in turn, can enhance the findability, interoperability, and reuse of open datasets on the Web---especially in the context of RDF datasets---contributing to a more open and interconnected data ecosystem.

\textbf{Reusability.}
Our test collection has been made openly available on Zenodo, while the source code for the adapted baseline methods has been released on GitHub. Both repositories are accompanied by comprehensive documentation, usage examples, and ready-to-run scripts to facilitate seamless adoption.

\textbf{Maintenance Plan.}
Although our current LLM-annotated training cases have demonstrated promising utility in the experiments, we plan to continuously supplement these with additional manually annotated cases. To encourage a broader adoption of DSEBench, we will also submit proposals to host a dedicated challenge at ISWC or a shared task at an IR conference focused on DSE and/or ExDSE. Additionally, while the present version of DSEBench builds on the NTCIR test collection, we intend to extend its coverage by adapting other relevant test collections---especially those centered on RDF datasets~\cite{ACORDAR1,ACORDAR2}---using a similar methodology in future iterations.

\textbf{Limitations.}
DSEBench has the following limitations, which we plan to address in future work.
First, the queries in DSEBench are each associated with only a single target dataset. A more general evaluation setting for DSE should accommodate queries with multiple target datasets. However, constructing such test cases requires identifying a larger set of datasets known to be relevant to each query, which would reduce the total number of test cases that can be feasibly built under current constraints.
Second, our pooling methods simply expand the original query by concatenating it with the fields of the target datasets to retrieve. Given the dual nature of DSE, it may be necessary to assess query relevance and target similarity separately---an objective we propose for future research into dedicated methods.
Third, our annotations for target similarity rely on a general, context-agnostic notion of similarity between datasets. In practice, the criteria for judging similarity in DSE may vary depending on specific user contexts, reflecting more nuanced information needs and evaluation scenarios.

\subsubsection{Use of Generative AI.}
In Section~\ref{sec:llmann}, we detail the use of GLM for data annotation. The paper text was also refined with the assistance of DeepSeek.

\subsubsection{Acknowledgements.}
This work was supported by the National Science and Technology Innovation 2030 Major Program (2025ZD0544900).

\clearpage

%
%
%
\bibliographystyle{splncs04}
\bibliography{main}
%




\end{document}